\documentclass[12pt,preprint]{aastex}
\usepackage{graphicx}
\usepackage{amssymb}
\usepackage{lscape}

\shorttitle{Outflows from Massive Protostars}

\shortauthors{Klaassen \& Wilson}

\title{Outflow and Infall in a Sample of Massive Star Forming Regions II: Large Scale Kinematics}

\author{P. D. Klaassen, \& C. D. Wilson}
\affil{Dept. of Physics and Astronomy, McMaster University, Hamilton, ON, Canada}
\email{klaassp@physics.mcmaster.ca}

\begin{document}

\begin{abstract}
We present maps of seven sources selected from \citet{KW07a} in SiO (J=8-7) and HCO$^+$ and H$^{13}$CO$^+$ (J=4-3) which were obtained using HARP-B on the James Clerk Maxwell Telescope. We find that four out of our seven sources have infall signatures based on their HCO$^+$ emission profiles.  From our maps, we have determined the extent of both the outflowing and infalling regions towards these sources, and constrained the amount of infalling and outflowing mass as well as the mass infall rate for each massive star forming region. From our SiO observations, we estimate the source luminosity required to shock the surroundings of these massive star forming regions and find luminosities similar to those of the HII regions themselves. We find that the ratio between our infall and outflow masses is less than one, suggesting high mass entrainment rates in the molecular outflows.  We also find that the large scale molecular infall rate towards G10.6-0.4 is comparable to the small scale molecular infall rate found in previous studies. 
\end{abstract}


\section{Introduction}
\label{sec:intro}

In the last 20 years, the processes involved in the formation of solar type stars have been well characterized.  An evolutionary sequence  has been developed in which a prestellar core \citep[i.e.][]{Andre00} begins to collapse under its own gravity and forms a Class 0 object \citep[i.e.][]{Andre93} with a large disk of material surrounding it.  This accreting protostar begins to produce an observable molecular outflow which releases the build up of angular momentum \citep[i.e.][]{Bach96,Arce07} and evolves into the Class 1 stage.  Later, as the protostar becomes more revealed, accretion slows and a pre-main-sequence star forms.  It then undergoes quasi-static contraction until core hydrogen burning commences and the star reaches the main sequence \citep[i.e.][]{Shu87}. These processes have been well studied and characterized based in part on the high resolution observations possible towards the closest examples of low mass star forming regions.

The initial mass function (IMF) of stars \citep[i.e.][]{Kroupa01} shows that most of the mass contained in stars is within the lower mass stars, and that massive stars do not form nearly as often as their lower mass counterparts.  There are a number of nearby low mass star forming regions (i.e. Taurus, Ophiucus, Chameleon, and Serpens are well studied nearby examples) in which there is no massive star formation. High mass stars do not, in general, form in isolation, but together with low mass stars \citep[i.e. Orion, see][]{Riddick07}. The high densities required for massive star formation also mean that more of them form in the molecular ring. This means that statistically, the closest examples of regions forming high mass stars are further away, and are also forming clusters of stars. Both distance and clustering can cause confusion when attempting to determine the dynamics in a region.

Add to the large average distances and source confusion the higher energies involved in the formation of massive stars and the presence of HII regions surrounding the more evolved massive protostars, and it becomes clear why high mass star formation is not nearly as well understood as lower mass star formation.  If we assume that massive stars form in a manner similar to their lower mass counterparts \citep[i.e.][]{Mckee2003}, it is possible to suggest an evolutionary sequence starting from an Infrared Dark Cloud \citep[i.e.][]{Pillai06} with a deeply embedded protostar which then begins to heat its surroundings and creates a hot core \citep[i.e.][]{Kurtz00}.  The protostar then begins burning hydrogen brightly enough to ionize its surroundings, forming a hypercompact HII region \citep[][]{Keto03}, which continues to expand into an ultracompact HII region \citep[][]{WC89} as the protostar evolves towards and up the main sequence \citep{B+S05,KW06}. There are many observable phenomena, such as large scale infall and outflows, which appear to be common between both low and high mass star formation, which suggests that perhaps high mass stars may form via the accretion of matter onto a single object in a manner similar to low mass stars.  However, for high mass star formation, there are still a number of questions to be answered. For the large scale phenomena observable with a single dish telescope, we still do not know  how large an area the outflowing gas shocks, how much energy is required to shock these regions,  how large the infall region is,  or how much mass from the protostellar core undergoes infall onto the central massive object.

Here, we present follow-up observations to \citet[][hereafter Paper 1]{KW07a} with which we address the questions posed above for seven high mass star forming regions mapped using the James Clerk Maxwell Telescope\footnote{The James Clerk Maxwell Telescope is operated by The Joint Astronomy Centre on behalf of the Science and Technology Facilities Council of the United Kingdom, the Netherlands Organisation for Scientific Research, and the National Research Council of Canada.}. In Paper 1, we obtained single pointing observations towards 23 massive star forming regions previously known to have HII regions \citep{WC89,KCW94} and molecular outflows as traced by CS \citep{P97,S03} or CO \citep{Hunter97}.  We were able to show that, of our observed high mass star forming regions with ongoing outflow activity (14 sources, as traced by SiO emission), half (seven) showed evidence for large scale infall motion as well (as traced by HCO$^+$ line asymmetries).  We suggested that the sources with infall and outflow detections were still actively accreting material (possibly in an ionized form, like that discussed in \citet{Keto07},  while the other half may have already finished accreting \citep[like the source discussed in][]{msc}.  However, with single pointing observations, we could not determine the sizes of the infalling or shocked (outflowing) regions, and therefore could not discuss properties such as the extents of the infall and outflow regions, whether these phenomena were beam diluted, or how their sizes compared. In this paper, we present maps of seven sources in the same tracers as Paper 1 in order to understand better the nature of the large scale infall and outflow towards massive star forming regions.

For this paper, we chose to map the SiO (J=8-7) and HCO$^+$/H$^{13}$CO$^+$ (J=4-3) emission towards seven massive star forming regions.  Six sources were chose from Paper 1, and were chosen to be representative of the survey as a whole. Two sources (G29.96-0.02 and G5.89-0.39) showed only outflow signatures, while two showed both infall and outflow signatures (G10.47+0.03 and G19.61-0.23), and one source (K3-50A) showed only a weak outflow signature. It was unclear whether the emission towards the sixth source (G20.08-0.14) was due to one cloud (as suggested by the single peaked SiO line profile) or two (as suggested by the double peaked HCO$^+$ and H$^{13}$CO$^+$ line profiles).   The G20.08 region was previously known to have two ultracompact HII regions in close proximity \citep[i.e.][]{Sewilo04}, although the two HII regions are separated by more than a single JCMT beam.  To this list of six sources, we added one more source, G10.6-0.4.  This source was added to determine whether the ionized and molecular accretion observed on small scales in this region \citep[i.e.][]{KW06,Sollins05} has a corresponding larger scale molecular infall signature.  G10.6-0.4 will act as a good test for determining whether our large scale results reflect the smaller scale dynamics, since the small scale structures are fairly well understood for this region. From this point forward, all source names based on Galactic coordinates will be truncated to latitudes only.

In Section \ref{sec:observations} we present our observations and in Section \ref{sec:results} we determine outflow and infall properties based on these observations.  Our results are then discussed in Section \ref{sec:discussion}, and we summarize in Section \ref{sec:conclusions}

\section{Observations}
\label{sec:observations}

Observations of SiO (J=8-7), HCO$^+$ and H$^{13}$CO$^+$ (J=4-3) were obtained at the JCMT towards G10.47, G10.62, G19.61, G20.08, G29.96 and K3-50A in April and August 2007 as project M06BC08 along with HCO$^+$ observations of G5.89.  SiO (J=8-7) and H$^{13}$CO$^+$ were observed simultaneously towards G5.89-0.39 as part of project M03AC13 \citep[see][for a description of these observations]{msc} and taken from the JCMT archive for this study.  The archive map is smaller than our new maps; however, both the SiO and H$^{13}$CO$^+$ emission falls below the rms limits within the edges of this map.  

The spectral resolution of the new datasets is 0.42 km s$^{-1}$, the observations are spaced at 7$''$ to Nyquist sample the 14$''$ full width of the JCMT beam at these frequencies, and the main beam efficiency was 0.62.  Map center positions and local standard of rest velocities are given for our six original sources in Paper 1. For G10.6, these properties are as follows: \mbox{$\alpha$ = 18:10:27.8}, \mbox{$\delta$ = -19:56:04}, and \mbox{V$_{\rm LSR}$ = 3 km s$^{-1}$}. For reasons presented in Section \ref{sec:G1047}, in this paper we adopt a distance of 5.8 kpc \citep[i.e.][]{Gibb04} for G10.37, which is much closer than the value of 12 kpc \citep[i.e.][]{S03} used in Paper 1.

The 2007 observations were obtained using HARP-B and the ACSIS autororrelator system in either grid position switching or jiggle position switching mode.  Both methods provided similar noise limits in similar times.  The noise levels in each map are given in the respective figure captions and were generally consistent across each map, staying within approximately 30\% of a mean value when receivers with bad noise levels were removed from the maps.  These observations were reduced using the Joint Astronomy Center Starlink software.  The 2003 observations were obtained with RxB in raster mapping mode using the DAS autocorrelator system and reduced using SPECX. First order linear baselines were removed from each pixel within each map, and the reduced maps were exported to the MIRIAD format for analysis.

\section{Results}
\label{sec:results}

SiO and HCO$^+$ were chosen for this study because they trace outflow and infall, respectively, in star forming regions on all mass scales.  Once each map was properly calibrated and exported into MIRIAD, the peak brightness temperature and line full width at zero power (FWZP) were determined from the spectrum at the map pointing center.  The uncertainty in the integrated intensity for each line was determined using $\Delta I=$T$_{\rm rms}\Delta\nu($N$_{\rm chan})^{1/2}$ where T$_{\rm rms}$ is the single channel rms uncertainty, $\Delta\nu$ is the velocity resolution of our observations, and N$_{\rm chan}$ is the number of channels used to cover the full width of the line (FWZP).  Note that for sources with self absorbed HCO$^+$ profiles, the peak line temperature is an underestimate of the source brightness temperature.  

In our maps, we find that the HCO$^+$ emission towards each source is extended, as is the H$^{13}$CO$^+$ emission; however, H$^{13}$CO$^+$ does not reach the same spatial scales as the 77 times more abundant  (Wilson \& Rood, 1994) HCO$^+$. In only five of our seven sources is the SiO emission at least marginally resolved.  The deconvolved source size was determined using $D_{\rm sou} = \sqrt{D_{\rm obs}^2-D_{\rm beam}^2}$ where the longest axis of the observed region full width at half maximum diameter ($D_{\rm obs}$) was used to compare to the beam diameter ($D_{\rm beam}$).  Towards the two unresolved sources (K3-50A and G19.61), the deconvolved diameter of the emitting region is less than 50\% of the JCMT primary beam, while the deconvolved diameter is greater than 70\% of the JCMT primary beam towards our other sources. Our SiO detection is weak in K3-50A (the line peak was only detected at 3.5-4$\sigma$) and was insufficient for detecting SiO beyond the central pixel.  This was not the case for G19.61 (the line peak was detected at 7$\sigma$), yet the SiO emission in this source was also unresolved. Thus, if the shocked region in G19.61 were as extended as our JCMT beam, we would have detected the extended SiO emission.

\begin{figure}
\begin{center}
\vspace*{7cm}
\includegraphics{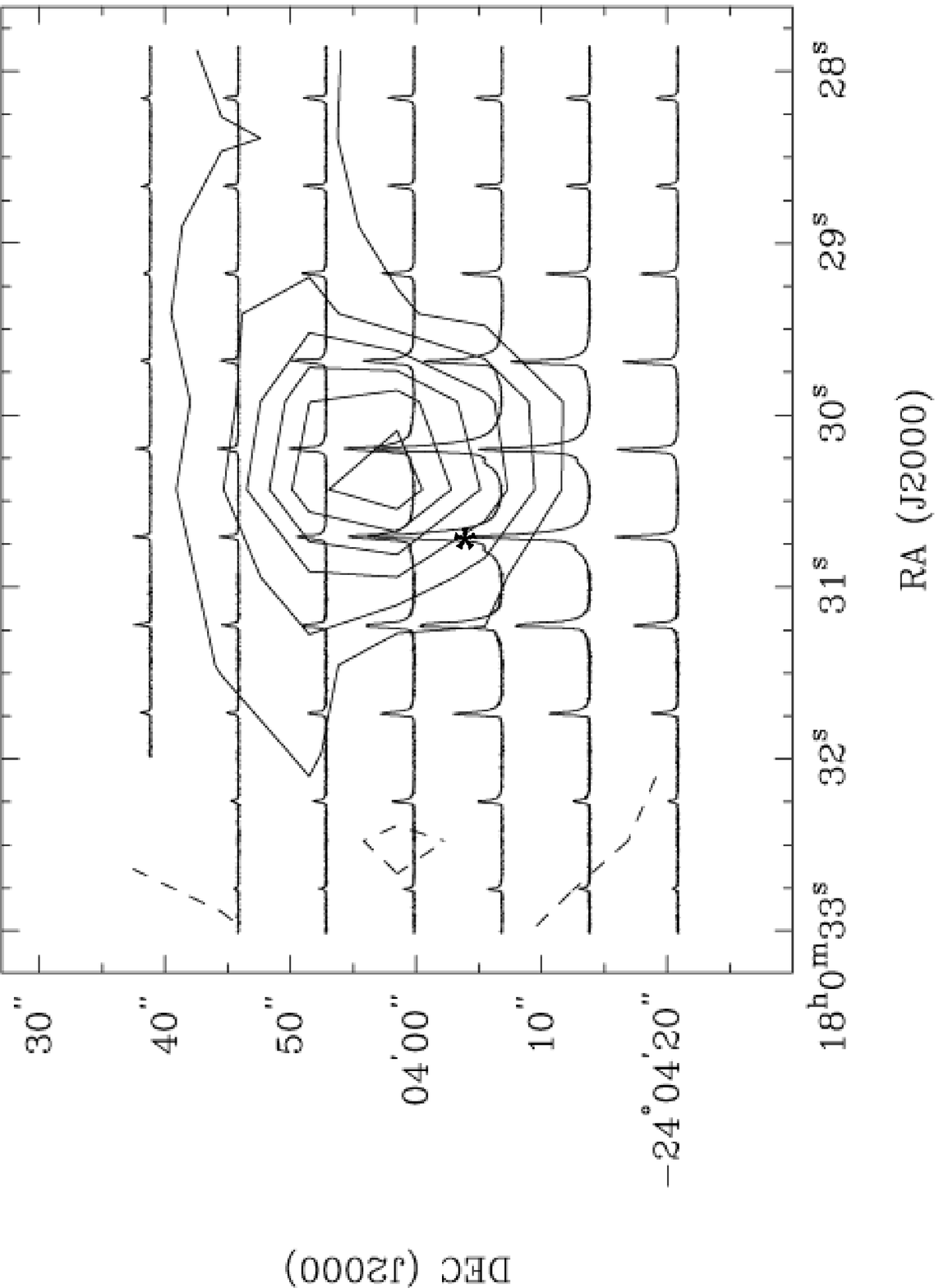}
\caption{{\bf G5.89-0.39} The solid contours represent integrated SiO emission towards G5.89. They start at 3$\sigma$ (9.68 K km s$^{-1}$) and increase in steps of 3$\sigma$. The dashed contour is the 3$\sigma$ (5.23 K km s$^{-1}$) contour of H$^{13}$CO$^+$ emission while the spectra represent the HCO$^+$ emission where the integrated intensity is above 10$\sigma$ (14.50 K km s$^{-1}$) from -50 to 80 km s$^{-1}$.}
\label{fig:G5map}
\end{center}
\end{figure}

\begin{figure}
\begin{center}
\vspace*{7cm}
\includegraphics{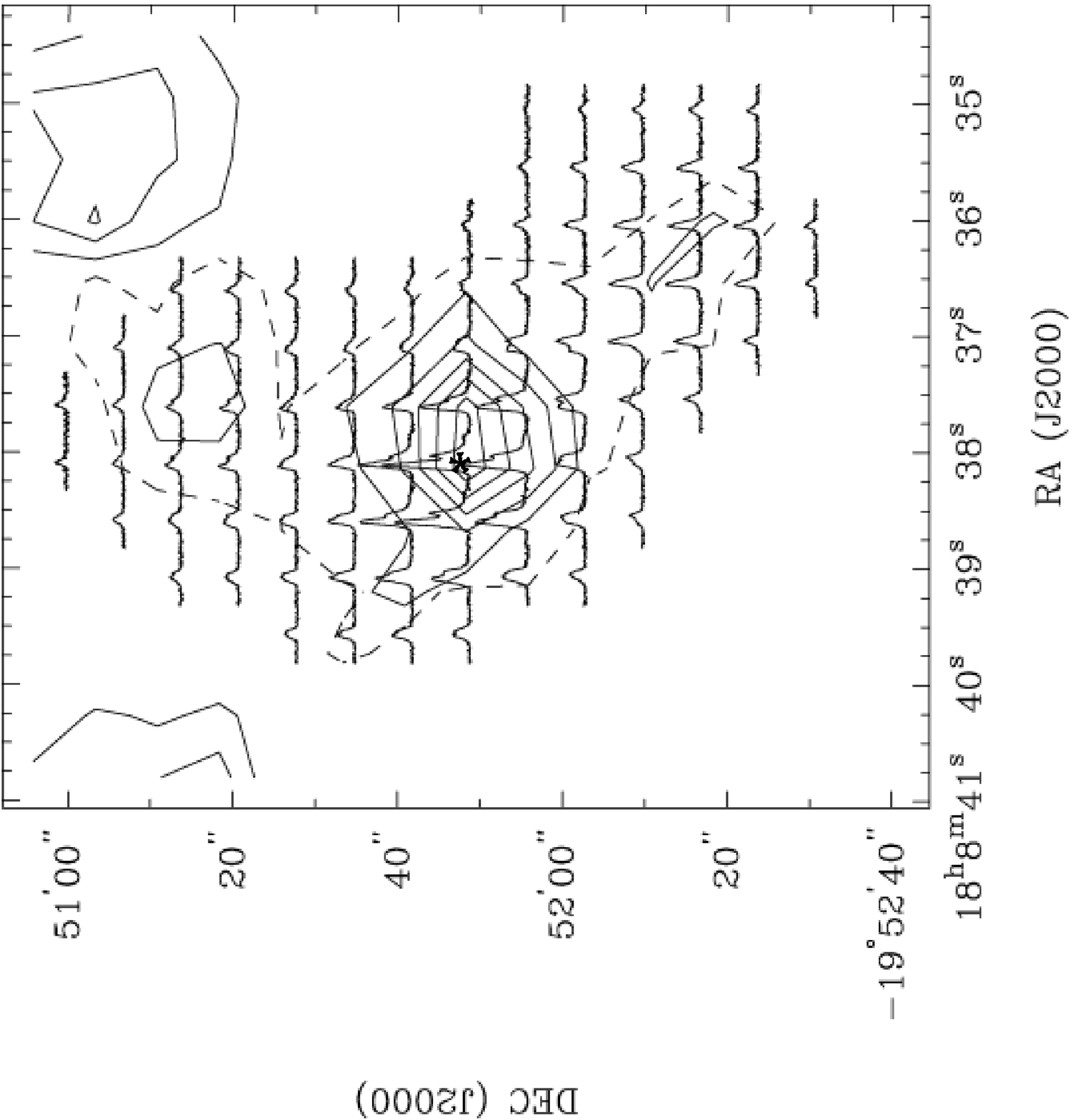}
\caption{{\bf G10.47+0.03} The solid contours represent integrated SiO emission towards G10.47. They start at 3$\sigma$ (1.14 K km s$^{-1}$) and increase in steps of 3$\sigma$. The dashed contour is the 3$\sigma$ (0.75 K km s$^{-1}$) contour of H$^{13}$CO$^+$ emission while the spectra represent the HCO$^+$ emission where the integrated intensity is above 10$\sigma$ (10.00 K km s$^{-1}$) from 45 to 90 km s$^{-1}$.}
\label{fig:G1047map}
\end{center}
\end{figure}

\begin{figure}
\begin{center}
\vspace*{7cm}
\includegraphics{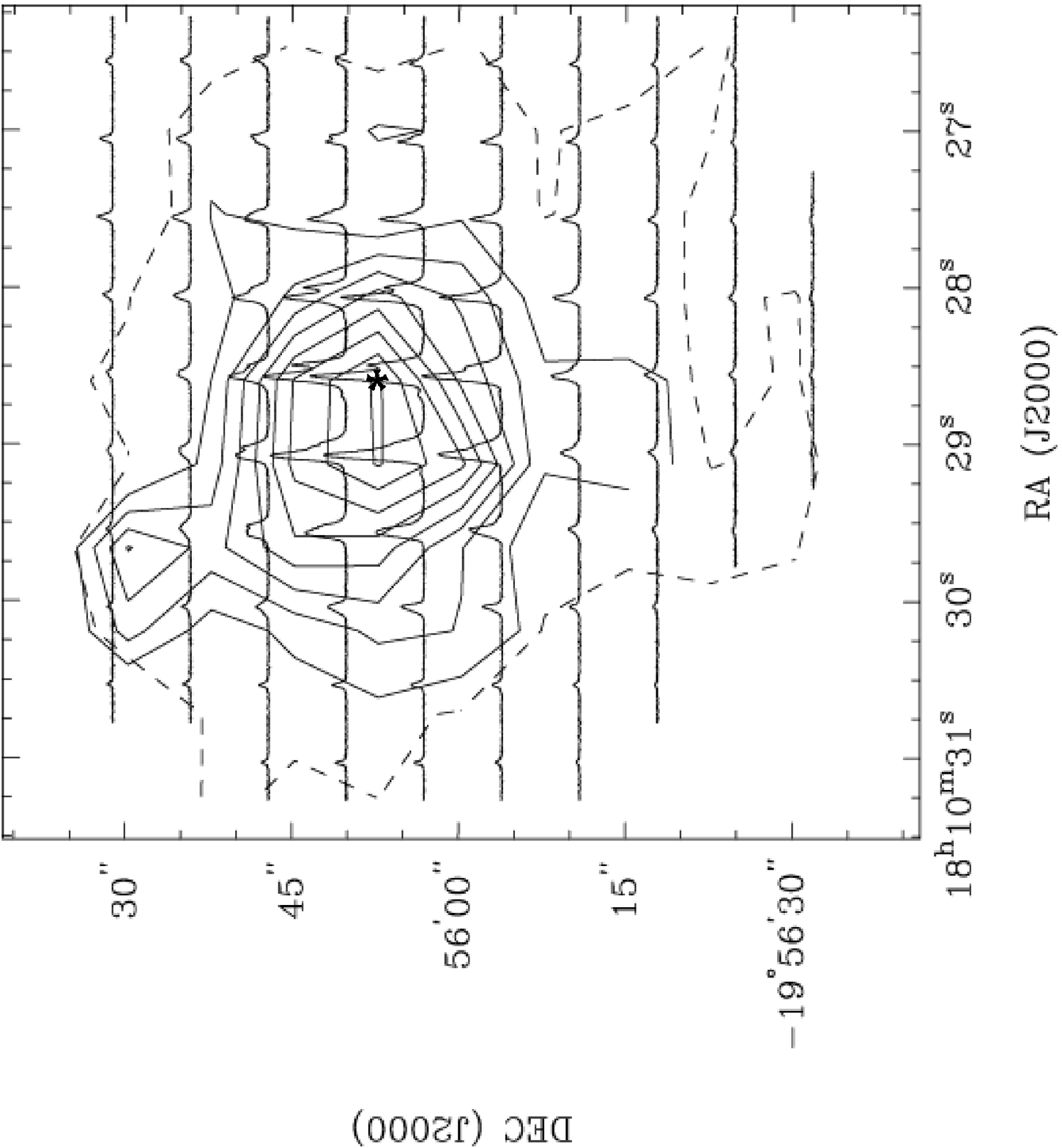}
\caption{{\bf G10.62-0.38} The solid contours represent integrated SiO emission towards G10.6. They start at 3$\sigma$ (1.01 K km s$^{-1}$) and increase in steps of 3$\sigma$. The dashed contour is the 3$\sigma$ (1.13 K km s$^{-1}$) contour of H$^{13}$CO$^+$ emission while the spectra represent the HCO$^+$ emission where the integrated intensity is above 10$\sigma$ (8.55 K km s$^{-1}$) from -30 to 30 km s$^{-1}$.}
\label{fig:G106map}
\end{center}
\end{figure}

\begin{figure}
\begin{center}
\vspace*{7cm}
\includegraphics{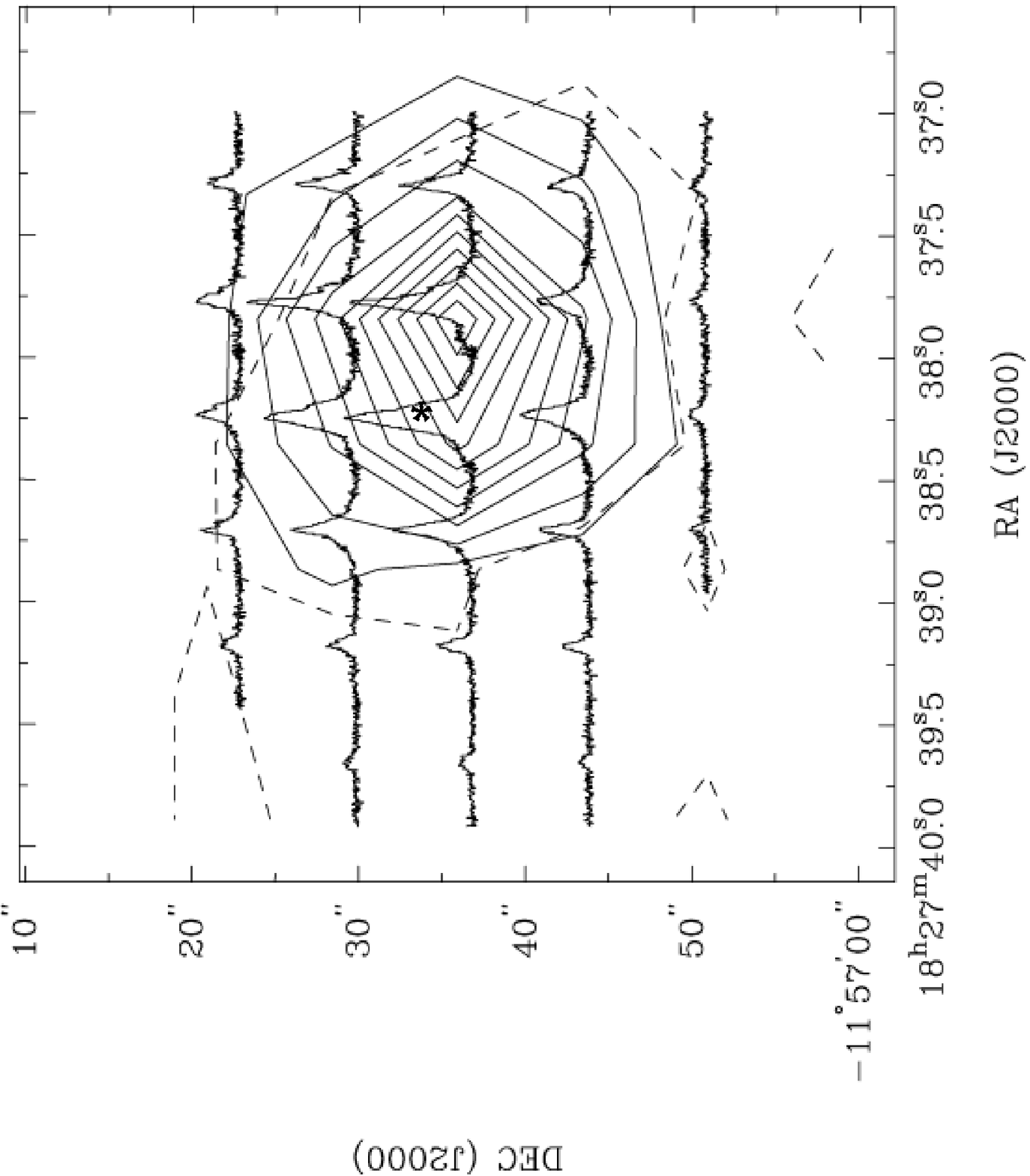}
\caption{{\bf G19.61-0.23} The solid contours represent integrated SiO emission towards G19.61. They start at 3$\sigma$ (1.35 K km s$^{-1}$) and increase in steps of 3$\sigma$. The dashed contour is the 3$\sigma$ (0.9 K km s$^{-1}$) contour of H$^{13}$CO$^+$ emission while the spectra represent the HCO$^+$ emission where the integrated intensity is above 10$\sigma$ (8.39 K km s$^{-1}$) from 10 to 80 km s$^{-1}$.}
\label{fig:G19map}
\end{center}
\end{figure}

\begin{figure}
\begin{center}
\vspace*{7cm}
\includegraphics{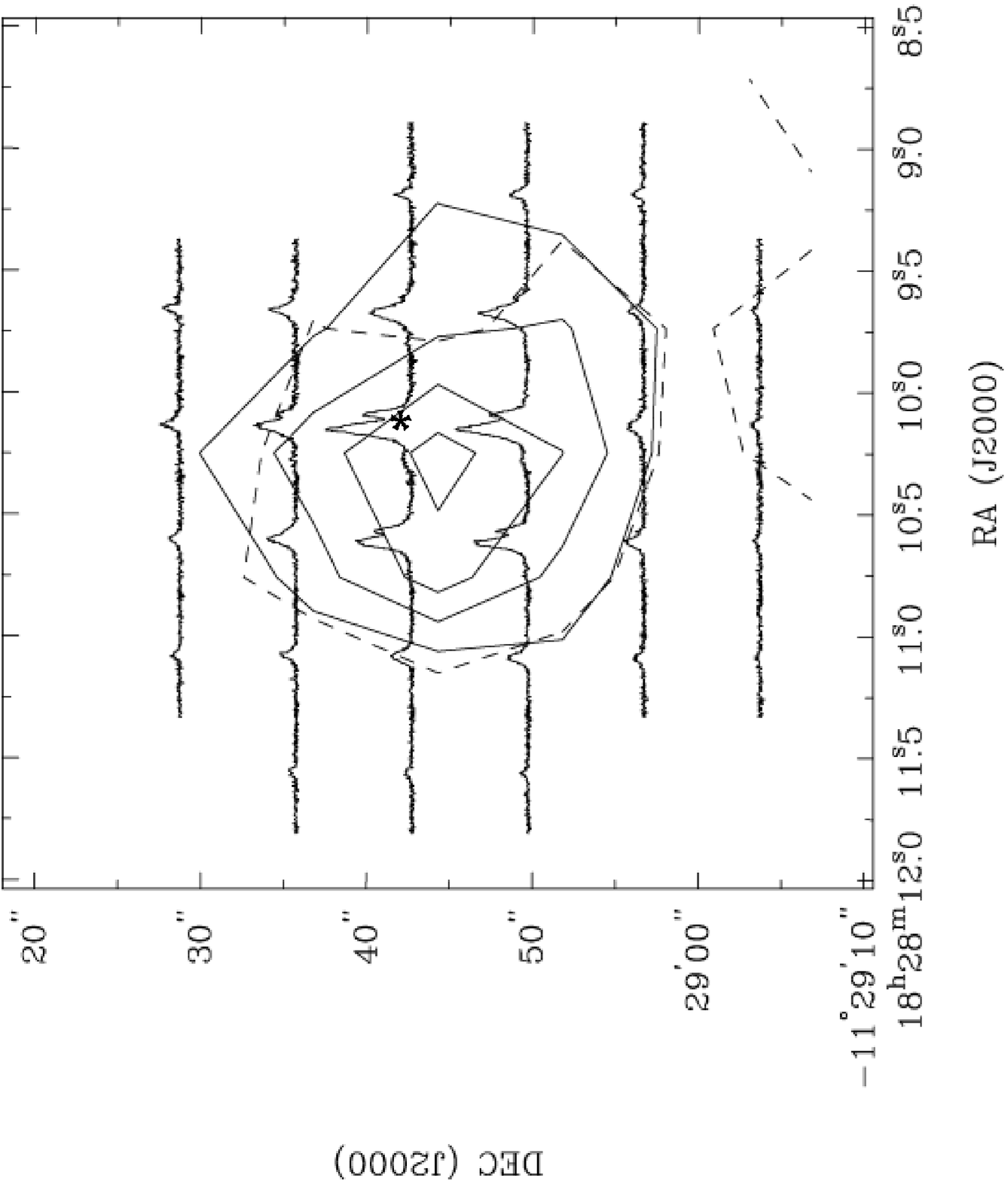}
\caption{{\bf G20.08-0.14} The solid contours represent integrated SiO emission towards G20.08. They start at 3$\sigma$ (1.10 K km s$^{-1}$) and increase in steps of 3$\sigma$. The dashed contour is the 3$\sigma$ (0.90 K km s$^{-1}$) contour of H$^{13}$CO$^+$ emission while the spectra represent the HCO$^+$ emission where the integrated intensity is above 10$\sigma$ (8.23 K km s$^{-1}$) from 10 to 80 km s$^{-1}$.}
\label{fig:G20map}
\end{center}
\end{figure}

\begin{figure}
\begin{center}
\vspace*{7cm}
\includegraphics{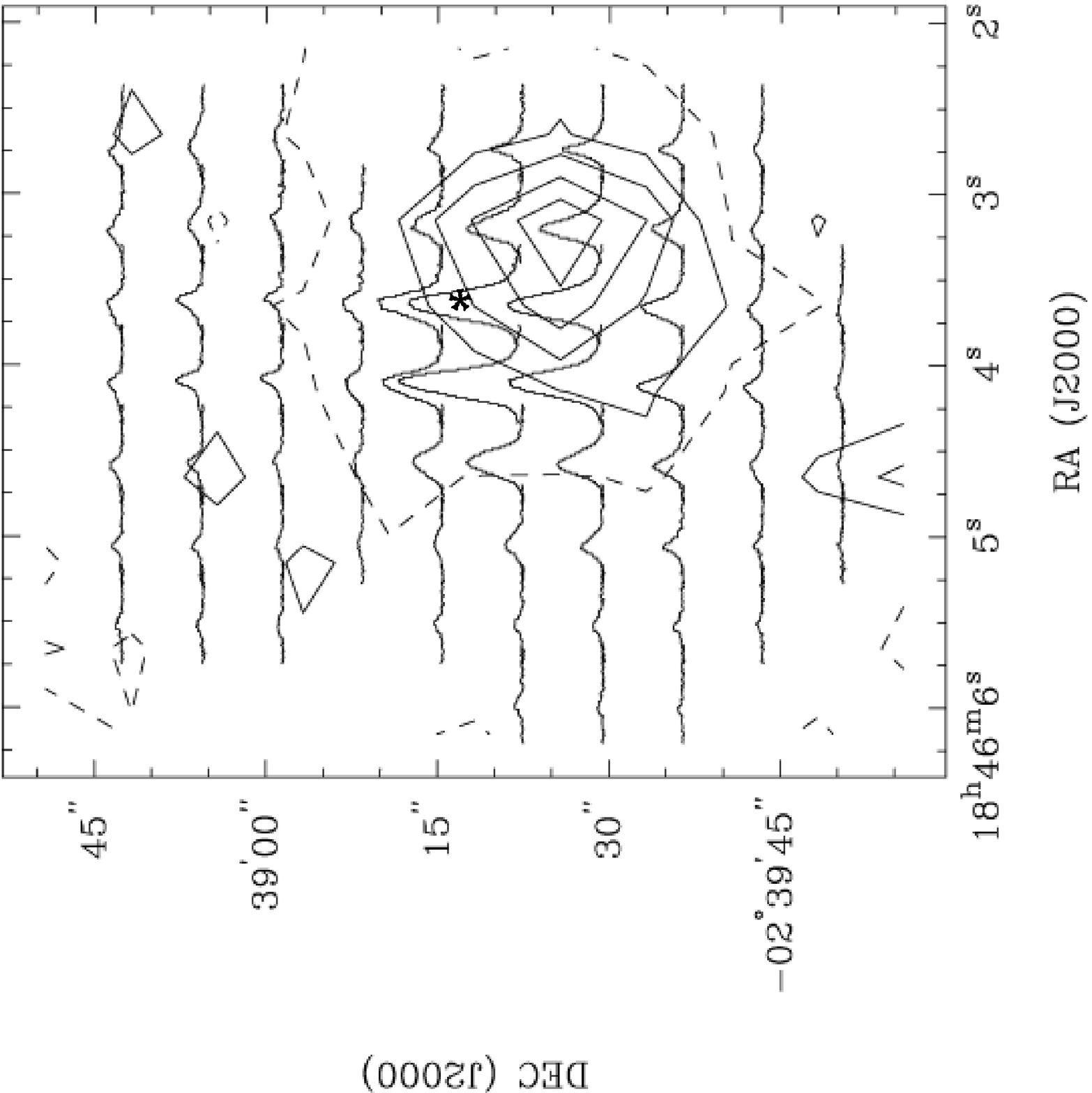}
\caption{{\bf G29.96-0.02} The solid contours represent integrated SiO emission towards G29.96. They start at 3$\sigma$ (1.53 K km s$^{-1}$) and increase in steps of 3$\sigma$. The dashed contour is the 3$\sigma$ (0.76 K km s$^{-1}$) contour of H$^{13}$CO$^+$ emission while the spectra represent the HCO$^+$ emission where the integrated intensity is above 10$\sigma$ (6.3 K km s$^{-1}$) from 90 to 110 km s$^{-1}$.}
\label{fig:G29map}
\end{center}
\end{figure}

\begin{figure}
\begin{center}
\vspace*{7cm}
\includegraphics{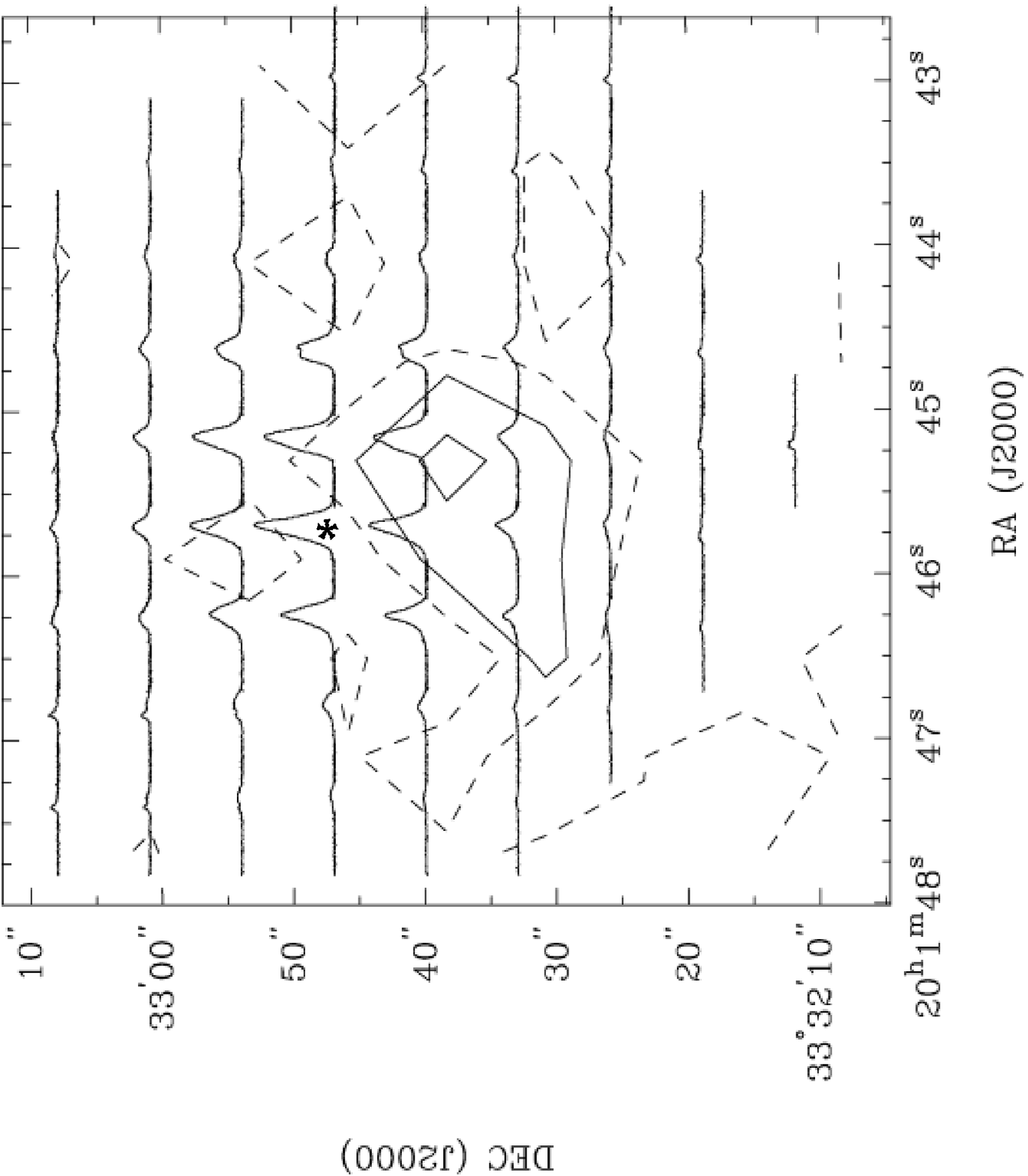}
\caption{{\bf K3-50A} The solid contours represent integrated SiO emission towards K3-50A. They start at 3$\sigma$ (1.15 K km s$^{-1}$) and increase in steps of 3$\sigma$. The dashed contour is the 3$\sigma$ (1.00 K km s$^{-1}$) contour of H$^{13}$CO$^+$ emission while the spectra represent the HCO$^+$ emission where the integrated intensity is above 10$\sigma$ (5.97 K km s$^{-1}$) from -50 to 0 km s$^{-1}$.}
\label{fig:K3-50map}
\end{center}
\end{figure}

Emission maps for each source are shown in Figures \ref{fig:G5map}-\ref{fig:K3-50map}. In each case, HCO$^+$ emission spectra are shown for pixels with HCO$^+$ integrated intensities which are 10$\sigma$ or greater. For each tracer, the intensity scale for the HCO$^+$ spectra is constant for a given source, and the peak HCO$^+$ intensity for each source is listed in Table \ref{tab:hco_prop}.  The dashed contour in each plot represents the 3$\sigma$ emission region for H$^{13}$CO$^+$ and the solid contours represent SiO emission, starting at 3$\sigma$ and increasing in intervals of 3$\sigma$.  The 3$\sigma$ rms uncertainties in the integrated intensities are given in each figure caption. The star in each figure marks the position of the peak in the HCO$^+$ integrated intensity.  In all cases except K3-50A and G29.96, the positions of the HCO$^+$ peak is within the same beam as the SiO peak. For the two remaining sources, this offset may be due to pointing errors between the two maps.

Using the FWZP from the central pixel, the integrated intensity for each species across each map was determined.   The rms uncertainty in the integrated intensity was determined using the same method as presented in Paper 1.  For each tracer, the line is broadest in the map center, and so using the same FWZP across the entire map captures all of the emission from that transition. The integrated intensities averaged over each map in SiO and HCO$^+$ are shown in Tables \ref{tab:sio_prop} and \ref{tab:hco_prop}, respectively.

The integrated intensities in SiO and HCO$^+$ were used to find the column density of gas emitting across the map in that tracer.  For optically thin gas, the column density in the higher energy level is proportional to the integrated intensity of the line, with a scaling constant proportional to the frequency of the transition and the Einstein coefficient for that transition \citep[see Equation 1 of Paper 1, or][]{Tielens}. Assuming local thermodynamic equilibrium and an ambient temperature of 44 K (as was done in Paper 1), we can use the partition function to determine the column density of optically thin gas in each tracer \citep[see for instance Equation 14.38 of][]{RW04}.  The average column densities of SiO and HCO$^+$ are presented in Tables \ref{tab:sio_prop} and \ref{tab:hco_prop}.  The values for the H$^{13}$CO$^+$ column density are not presented here, but were found to be only 4-22 times less than those for HCO$^+$.  That we do not see a larger difference in the column densities, given the abundance ratio of 77 between the isotopologues, suggests that we are not seeing all of the HCO$^+$ emission because the line is optically thick. Thus, our derived column densities are lower limits to the true HCO$^+$ column densities due to optical depth effects. This trend was also seen in Paper 1, and supports the claim that the double peaked HCO$^+$ profiles towards four of our sources are due to self absorption of the optically thick line, and not multiple clouds along the line of sight.

\subsection{Outflow properties derived from SiO observations}
\label{subsec:sio}

In an environment which has recently been shocked, the abundance of SiO can be enhanced up to a factor of 10$^{-8}$ (from  dark cloud abundances of approximately 10$^{-11}$) with respect to molecular hydrogen, due to silicon being liberated from dust grains and combining with oxygen to form SiO \citep[i.e.][]{Caselli97,Schilke97}.  After approximately 10$^4$ years, SiO depletes out of the gas phase again and the abundance drops \citep[i.e.][]{Pineau97}.  Because of the short enhancement lifetime, SiO is generally only detectable in regions which have recently been shocked. Where we detect SiO (J=4-3) we suggest there is ongoing outflow activity.

We present the column density of SiO towards each source in Table \ref{tab:sio_prop}. We can divide this value by the abundance of SiO with respect to H$_2$ in order to determine the column density of shocked gas in each region. These values are also presented in Table \ref{tab:sio_prop}.  We used the same methods as described in Paper 1 to calculate the abundance of SiO towards G10.6 \mbox{(log[$N_{\rm SiO}$] + log[$X_{\rm CS}$] - log[$N_{\rm CS}$] = log[$X_{\rm SiO}$] = -9.92)} and we have assumed a constant SiO abundance across each map. We note that the SiO column densities listed in Paper 1 are over estimated by a factor of 3.8, and thus the logarithmic SiO abundances listed in that paper are corrected in Table \ref{tab:sio_prop}.  Because we have adopted a new distance for G10.47, we have had in addition to modify our original abundance estimate from Paper 1.  The CS column density quoted in \citet{P97} appears to be distance independent; however, the CS abundance presented in \citet{S03} is dependent on the assumed distance to the source (see their Equations 7 and 12).  Thus, our SiO abundance has changed from -9.53 in Paper 1 to -10.42 when the new distance and revised column density are taken into account. The SiO abundance determined for K3-50A in Paper 1 \mbox{(log[$X_{\rm SiO}$] = -11.6)} is much closer to that found in dark clouds than in outflow regions.  The region in which SiO is emitting is unresolved by our observations and highly beam diluted. Thus, for our analysis, we have assumed an SiO abundance of -10.15, the average abundance from our other six sources.  Based on the abundance estimates presented in Table \ref{tab:sio_prop}, we have determined the total column of shocked material towards each source, from which we determined the mass of shocked gas because we  know the area of the emitting region (see below). For our sources which are unresolved in SiO, we have taken the radius of our primary beam to be the radius of the emitting region. 

Comparing the integrated intensities of our SiO emission in the central pixel of each map, where our signal to noise ratio is highest, to the models presented in \citet{Schilke97}, we suggest shock velocities of 20-22 km s$^{-1}$, and ambient densities of $n_o=10^6$ cm$^{-3}$. The ambient densities suggested here are consistent with those of \citet{P97} who determined densities and CS column densities for these regions using large velocity gradient models.  For each of our sources, we adopt a shock velocity of $v_s=21$ km s$^{-1}$. 

Because we have mapped the SiO emission towards seven sources, we can determine the radius of the region in which SiO is being emitted (R$_s$).  The method used to determine the deconvolved source size was described in Section \ref{sec:observations}, and the radius of the emitting region is presented in Table \ref{tab:sio_prop}. From this radius and the shock velocity determined above, we can constrain the ages of our outflows from simple kinematic arguments.  These approximate outflow ages are presented in Table \ref{tab:hco_prop} and are used later to constrain the amount of infalling material.

In order for the outflow shocks to reach their observed sizes given the mass of shocked gas, the powering sources for these outflows must produce enough energy to push through the ambient medium.  If we assume a steady wind is driving a spherical bubble of gas outward, we can use the model of \citet{Castor75} to determine the luminosity required to push the ambient gas.  Equation 6 of \citet{Castor75} can be re-expressed as:

\begin{equation}
L_o = R_s^2\rho_ov_s^3\left(\frac{154\pi}{125}\right)
\end{equation}

\noindent where $L_o = E_ot$ is the luminosity imparted by the wind, $R_s$ is the radius of the bubble, $v_s$ is the shock velocity, and related to the outflow age by $v_s=R_s/t$, and $\rho_o$ is the ambient density, and related to the ambient number density by the mass of molecular hydrogen ($\rho_o=\mu m_H n_o$). While much of this wind luminosity goes into accelerating the ambient medium, some of it is also transferred into thermal energy. From \citet{Basu99}, the thermal energy can be expressed as $E_{\rm th}=(5/11)L_o t$, or, given the value for $t$ above, $E_{\rm th}=(5/11)L_o R_s/v_s$.  Both the calculated wind luminosity and thermal energy required to power these wind driven shocks are listed in Table \ref{tab:sio_prop}.  These thermal energies are of approximately the same order of magnitude as the kinetic energies calculated for the CO outflows in a number of massive star forming regions \citep[see for instance][]{Wu04,KW07b}; similarly the wind luminosities are comparable to the luminosities of their HII regions \citep[see for instance][]{WC89}.

\subsection{Infall properties derived from HCO$^+$ observations}
\label{subsec:hco}

Many authors suggest that an asymmetric self absorbed line profile of an optically thick line tracer in which the blue emission peak is brighter than the red, and the optically thin tracer has a single line peak, can be indicative of infall \citep[i.e.][]{Mardones97,Gregersen97,Fuller05}.  Here, our optically thick line is HCO$^+$ (J=4-3) and the optically thin isotopologue is H$^{13}$CO$^+$ (J=4-3).

From Paper 1, we expected to observe infall signatures in G10.47 and G19.61.  Infall is detected towards these two sources, as well as towards G20.08 from our original study.  In our single pointing observations, G20.08 was shown to have a double peaked line profile in both HCO$^+$ and H$^{13}$CO$^+$, which was suggested to be due to two overlapping sources.  However, in the more extended map, we see that H$^{13}$CO$^+$ becomes single peaked, suggesting that perhaps it is optically thick towards the center of the source and that mapping is required to resolve the ambiguity with respect to the chance super-position of line of sight clouds and self absorption of multiple isotopologues.  G10.6 was not in our original sample, but was added because it was known to have small scale infall motions. It too is seen to have a large scale infall signature.

In Paper 1, we suggested that beam dilution may have been a contributing factor as to why we did not detect infall in a number of sources. That we have detected extended infall in two distant ($\sim$ 6 kpc)  sources in this study suggests that if beam dilution of the signal is a problem, it only affects intrinsically smaller sources, and not intrinsically more distant ones, since our unresolved detections are for G19.61 and G20.08 at 4.5 and 4.1 kpc. 

Infall velocities were calculated for each self absorbed HCO$^+$ spectrum towards G10.47, G10.6, G19.61, and G20.08 using the models of \citet{Myers96} in the same way as described in Paper 1.  From our rms uncertainties in line temperatures and the spectral resolution of our observations, we determined the uncertainties associated with our derived infall velocities. We used these uncertainties to limit the regions in which we claim infall detections. Only if the derived infall velocity is more than three times our uncertainty for that spectrum do we claim to detect infall in that pixel. Towards G19.61 and G20.08, robust detections were made in one and three pixels (respectively) suggesting that the infall in both of these sources is unresolved. Towards G10.47 and G10.6 (both at approximately 6 kpc), robust infall detections were made in seven Nyquist sampled pixels per map, suggesting that the large scale infall in these regions is extended.  

Along the northern, blue shifted, outflow lobe in G10.47, the HCO$^+$ line profiles again become double peaked.  It is quite possible that there is a second instance of a collapsing region here since SiO is marginally detected (at the 3$\sigma$ level) in the middle of this northern extension.  The double peaked profiles do lead to infall velocities (in three pixels) which are greater than 3$\sigma$; however, we do not include this region in our infall analysis for a number of reasons.  The infall signature is not continuous from the central region to this northern region, which suggests that there are two distinct regions of infall.  This discontinuity would confuse  conclusions about the central infall region.  Similarly, we are interested in the infall activity surrounding HII regions, where we know massive stars are forming.  Neither the mid infrared nor radio maps of this region (\citet{Pascucci04} and \citet{WC89}, respectively) extend far enough to the north to determine whether there are HII regions in this northern extension. Archival SCUBA maps of this region at the same resolution as our observations \citep{scuba08} show emission which has the same extent and morphology as our HCO$^+$ observations. 

For each region in which infall was observed, we determined an average infall velocity by taking the mean of the infall velocities determined for each position in each map using the models of \citet{Myers96}. This averaging resulted in the velocities shown in Table \ref{tab:hco_prop}. We chose to use this method over creating an averaged spectrum and determining infall velocities from it. The latter method resulted in unreasonably  high infall rates (i.e. 5 km s$^{-1}$ in G10.47), which give very large mass infall rates (0.22 M$_{\odot}$ yr$^{-1}$) and total infall masses ($\sim 2000$ M$_{\odot}$) assuming constant infall rates over the lifetime of the outflow. Averaging the spectra and then determining the infall velocity increased the infall velocity by increasing the full width at half maximum (FWHM) of the line. Since the infall velocity varies as the FWHM squared \citep[see Equation 9 of][]{Myers96}, this broadening leads to an overestimate of the infall velocity. Thus, for our unresolved sources, the calculated infall velocities may be upper limits, since there could be unresolved velocity gradients within the JCMT beam. Higher resolution imaging of these sources could decrease our calculated infall velocities.

Using the derived average infall velocities for each region, we determined mass infall rates for each sources using $\dot{M}=dM/dt\approx\rho Vv_{\rm in}/r = (4/3)\pi n_{o}\mu m_H r^2 v_{\rm in}$, where $\rho=\mu m_H n_{o}$ is the mass density, $v_{\rm in}$ is the infall velocity presented in Table \ref{tab:hco_prop}, and $V/r$ is the area of the infalling area (which we assume to be circular). Since the area in which we detected an infall signature was always similar to the area in which SiO was detected, we used the radius of the SiO emitting region as the radius of the infalling region.

From the radii of the SiO emitting regions and a shock velocity of 21 km s$^{-1}$, we were able to estimate outflow ages in Section \ref{subsec:sio}. Assuming infall and outflow are contemporaneous, these outflow ages can be used to constrain how much mass has undergone infall. These values are also given in Table \ref{tab:hco_prop}, and represent the total mass of infalling molecular gas, not just the mass of the infalling HCO$^+$.

\section{Discussion}
\label{sec:discussion}

It is interesting to note that the infall and outflow signatures in G19.61 are both unresolved, and the signatures in G20.08 are both marginally resolved. This comparison suggests that the shocked region, as traced by SiO, may have a similar physical size as the infall regions in these smaller sources.  Towards G10.6 and G10.47, the regions over which we have robustly detected extended infall, our deconvolved SiO emitting regions also appear to be on the same approximate spatial scales.  

Comparing infall mass and outflow mass, we note that our infall masses are lower than our outflow masses. This is contrary to the results of \citet{Beuther02}; however, their ratio was calculated for the protostellar jet, not the entrained molecular outflow. This comparison to \citet{Beuther02} suggests that more of the ambient mass is entrained into the large scale molecular outflow (produced by a protostellar jet) than falls in onto the forming stars \citep[i.e.][]{SC96}.

\subsection{G10.47: Extended outflow or chance super-position?}
\label{sec:G1047}

The large spatial extent of the HCO$^+$ emission in our map of G10.47 places it among the largest molecular outflows due to star formation in our Galaxy.  Some Herbig-Haro objects have been observed emanating from nearby lower mass star forming regions at distances of up to 12 pc from their powering stars \citep[i.e.][]{Bally02}; however in those cases, the associated molecular outflow is not nearly as wide as the case of G10.47, and those jets are not pushing through an ambient medium dense enough for massive stars to form (i.e. 10$^6$ cm$^{-3}$). In Paper 1, we used a distance estimate of 12 kpc to G10.47. Given the extent of HCO$^+$ gas ($\sim 2'$) in our current maps, this distance would give a linear scale of over 4 pc for a single outflow. Other authors have suggested a distance to this source of 5.8 kpc \citep[i.e.][]{Gibb04}. For the analysis presented here, we adopt this new distance, which changes the length of the outflow from 4 pc to 2 pc.  Given that this source is along a line of sight reasonably close to both the Galactic plane and Galactic center, it is not unreasonable to suggest that what appears to be an extremely large outflow is instead the super-position of a number of clouds along the line of sight.  However, based on the arguments presented below, we suggest that the gas kinematics and morphology are due to a large outflow.

\begin{figure}
\begin{center}
\vspace*{7cm}
\includegraphics{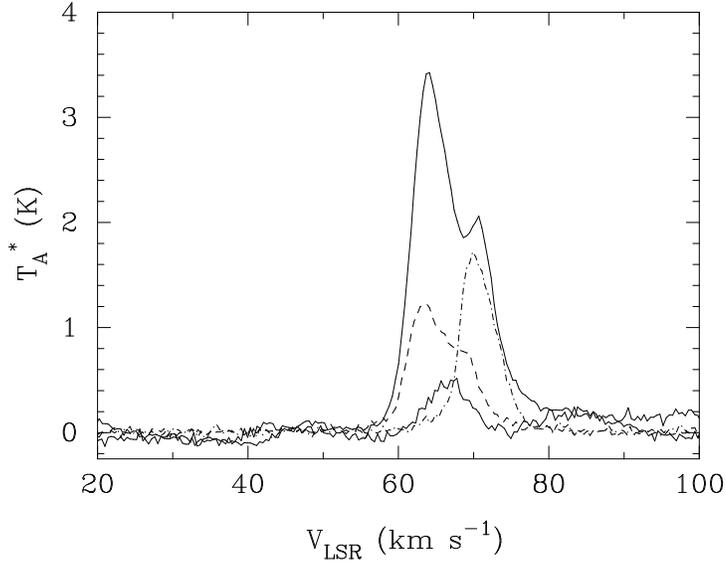}
\caption{Spectra from G10.47 averaged over 9 pixels each. The strongest (solid) line represents the HCO$^+$ emission towards the source center, while the dashed and dashed-dotted spectra represent the HCO$^+$ emission  towards the blue and red shifted regions (respectively). The least intense (solid) line represents the H$^{13}$CO$^+$ emission towards the same central region as the strong HCO$^+$ line.  Note that the central HCO$^+$ emission spectrum has a double peaked line profile due to self absorption, while the H$^{13}$CO$^+$ appears single peaked and optically thin.}
\label{fig:averaged_spectra}
\end{center}
\end{figure}

Examining the spectra of our HCO$^+$ map, we note that while the HCO$^+$ emission is extended, it is not present in all pixels, suggesting that we have captured all of the emission in this velocity range. Figure \ref{fig:averaged_spectra} shows the HCO$^+$ line profiles averaged over the blue and red shifted emission regions as well as the HCO$^+$ and H$^{13}$CO$^+$ line profiles averaged over the central nine pixels of our map.  The red and blue spectra, while having very little overlap between their profiles, do both fit within the width of the source center line profile.  The velocities at line peak for the blue and red spectra are -64 and -70 km s$^{-1}$ respectively, and match the velocities of the two peaks of the self absorbed, averaged HCO$^+$ central profile.  The single peak of the optically thin H$^{13}$CO$^+$ is at -67 km s$^{-1}$ as is the minimum of the HCO$^+$ self absorption feature.  This self-absorption lies directly between the peaks of the red and blue profiles, and is the same as the source rest velocity.  No other sources in the W31 region have rest velocities similar to the ones seen in the G10.47 region. 

The double peak in the HCO$^+$ spectrum is due to self absorption, not multiple components, since there is a single peak in the H$^{13}$CO$^+$ spectrum. The same is true for the CO isotopologues in this region. The BIMA observations of \citet{Gibb04} show one peak in C$^{18}$O, while the $^{13}$CO observations of \citet{Olmi96} with the IRAM Interferometer show two peaks indicative of self absorption.

\begin{figure}
\begin{center}
\vspace*{7cm}
\includegraphics{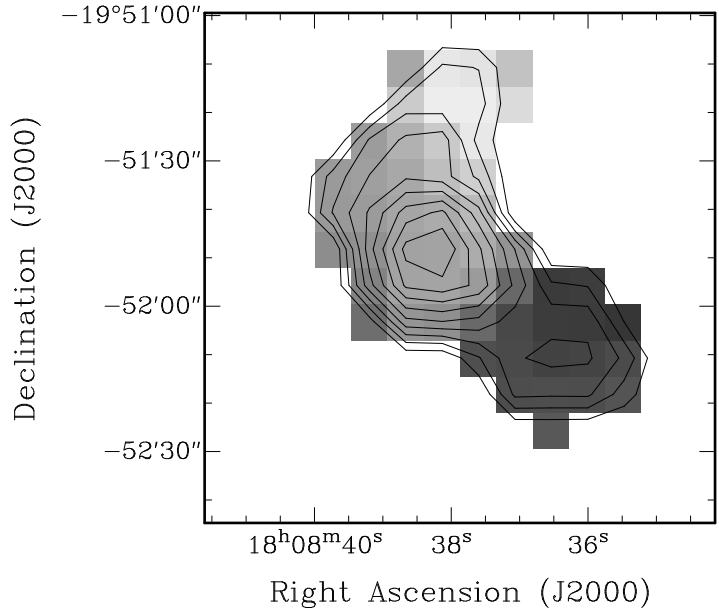}
\caption{First moment map of G10.47 (gray scale). The grayscale ranges from 63 to 72 K km s$^{-1}$. The velocity weighted integrated intensity shifts smoothly from the blue lobe to the central source to the red outflow lobe, suggesting that this is a single source with an outflow. The second moment map (black contours) shows that the highest velocity dispersion is centered on the location of the HII regions. The contours start at 10\% of the peak velocity dispersion (0.41 km s$^{-1}$) and increase in 10\% intervals.}
\label{fig:G1047_mom1}
\end{center}
\end{figure}

The first moment map of the HCO$^+$ emission (grayscale in Figure \ref{fig:G1047_mom1}) in this region also suggests a single outflow.  The observed smooth transition from blue to red along the source would be more likely to be observed in one source with a velocity gradient than three separate sources with discontinuous cloud boundaries.  The second moment map of this region (contours in Figure \ref{fig:G1047_mom1}) suggest that the largest velocity dispersion is towards the center, and continuously decreases towards the emission edges (outflow lobe ends), suggesting that most of the mass is centrally concentrated.

While we favor the outflow interpretation of this morphology, higher resolution, large scale observations of this source are necessary in order to resolve this ambiguity. This region contains four individual UCHII regions and what is likely a large scale outflow with a collimation factor ($R_{\rm maj}/R_{\rm min}$) of approximately 1.5, which may be even more highly collimated at higher resolution. Objects like G10.47, with large scale, fairly collimated outflow emanating from a cluster of HII regions pose interesting challenges for accretion models.

\subsection{Large and Small Scale Infall in G10.6}

As stated previously, G10.6 was added to our source list because it was previously known to show evidence for small scale infall, both in molecular gas \citep{Sollins05} and ionized gas \citep{KW06}.  These small scale infall signatures appear to be consistent with a single accretion flow which changes from being molecular to ionized as it passes the ionization boundary of the HII region.  While we cannot determine the orientation of the large scale infall from our observations, we note that our derived mass infall rate (0.03 M$_\odot$ yr$^{-1}$) is comparable to, but slightly larger than, that of \citet{Sollins05} (0.02 $M_{\odot}$ yr$^{-1}$).  This similarity may suggest that the large scale infall possibly maps to the small scale accretion flow; in addition, despite resolving the infall region, we may still be over estimating the infall velocity due to velocity gradients in our larger JCMT beam (see Section \ref{sec:hco}).

\subsection{How mapping has changed the results from Paper 1}

None of our basic conclusions drawn from our single pointing observations of SiO (i.e. column densities in the central pixel or outflow lifetimes) have changed due to mapping save for the infall detection in G20.08.  However, with respect to the outflowing gas, we have been able to place constraints on the energy required to shock these regions. With respect to infalling gas, we have placed better constraints on infall velocities and shown that the lack of infall signatures in some sources is not due to the signature being beam diluted out of observability at large distances.

We have found that one of our sources which was previously not thought to have an infall signature (G20.08) in fact does appear to have large scale infall motions. This discovery suggests that more accurate infall results are possible with maps of massive star forming environments than with single pointing observations.  Returning to the results of Paper 1, we determined that seven out of the 14 sources with ongoing outflow signatures had infall signatures as well.  These numbers can now be revised to eight out of 15 outflow sources with infall signatures when G10.6 is added to the source list.  Thus, we find that approximately half of the sources with ongoing outflows have infall signatures which we take to trace accretion.  Because our sources were selected based on the presence of HII regions, ongoing infall suggests that there is a phase of massive star formation in which the star is still actively accreting, but is also burning hydrogen brightly enough to ionize its surroundings \citep[see for instance the discussion in][]{KK08}.

\section{Conclusions}
\label{sec:conclusions}

We have found that just over half of the sources in our sample have large scale infall motions suggesting that the protostars powering the UCHII regions and massive outflows are still actively accreting material.  Where we have detected infall motions, they have been on quite large scales, suggesting that infall, as traced by HCO$^+$, is in general observable with single dish maps at the distances to massive star forming regions. Maps are required in order to capture all of the infalling gas since the infall signature can be extended, even at 6 kpc. In mapping these regions, we have been able to determine how much mass is falling into the central regions of our sample sources, which we would not have been able to determine with our single pointing observations. We find that when the infall is resolved, the infalling mass is greater than 30 M$_\odot$.  Our derived mass infall rate towards G10.6 is similar to the small scale rate found by \citet{Sollins05}, suggesting that we may be observing the same physical process, but on different spatial scales.

 In general, we have found that SiO is not very extended; however, since we mapped these regions in SiO, we were able to determine the radius of the emitting region, the mass of shocked gas and how much energy is required to shock the gas surrounding a massive protostar. We find these energies and luminosities to be similar to the kinetic energy imparted to the outflowing gas and the luminosities of the HII regions themselves.  We also find that the region over which we detect an infall signature and the region in which we detect SiO is quite similar, and find that in these regions, the total infall mass is less than the total outflow mass (M$_{\rm in}$/M$_{\rm out}\sim$  0.1 to 0.8) assuming the two processes are contemporaneous.  
 
\acknowledgements

We would like to acknowledge the support of the National Science and Engineering Research Council of Canada (NSERC). P. D. K would also like to thank J. Wadsley for helpful discussions during the preparation of this manuscript.

\begin{landscape}
\begin{deluxetable}{rccccccccccc}
\tablecolumns{12}
\tablewidth{0pc}
\tablecaption{SiO Observations and Derived Outflow Properties}
\tablehead{
\colhead{Source} & \colhead{Dist} & \colhead{T$_{\rm mb}$} & \colhead{FWZP\tablenotemark{a}}& \colhead{X[SiO]}&\colhead{$<\int T_{\rm mb}^*dv>$} & \colhead{$<$log[N$_{\rm SiO}$]$>$} &\colhead{$<$log[N$_{\rm H_2}$]$>$} & \colhead{Mass} & \colhead{Radius\tablenotemark{b}} & \colhead{L} &\colhead{E$_{\rm th}$}\\
	& \colhead{(kpc)} &\colhead{(K)}&\colhead{(km s$^{-1}$)} & & \colhead{(K km s$^{-1}$)} & \colhead{(cm$^{-2}$)} & \colhead{(cm$^{-2}$)} & \colhead{(M$_{\odot}$)} & \colhead{($\times10^{17}$ cm)} & \colhead{($\times10^{3}$ L$_\odot$)} & \colhead{($\times10^{48}$ erg)} 
}
\startdata
G5.89-0.39	&2.0	&2.37	&105	&-9.48	&4.38	&12.81	&22.29	&3.7	&1.9	&1.1	&0.18\\
G10.47+0.03	&5.8	&1.24	&30	&-10.42	&2.70	&12.60	&23.02	&184	&5.8	&10.3	&5.1\\
G10.6-0.4	&6.0	&1.14	&20	&-9.92	&8.74	&13.11	&23.03	&497	&9.4	&27.2	&21.6\\
G19.61-0.23	&4.5	&0.71	&45	&-10.32	&4.57	&12.83	&23.15	&162	&$<$4.7&$<$6.8&$<$2.7\\
G20.08-0.14	&4.1	&0.66	&30	&-11.06	&2.68	&12.60	&23.66	&374	&4.0	&4.8	&1.6\\
G29.96-0.02	&9.0	&0.63	&40	&-9.70	&5.14	&12.88	&22.58	&124	&7.9	&19.2	&12.8\\
K3-50A\tablenotemark{c}		&8.6	&0.48	&20	&-10.15	&3.15	&12.67	&22.82	&276	&$<$9.0&$<$24.7&$<$18.7
\enddata
\tablenotetext{a}{Full Width at Zero Power of the SiO line.}
\tablenotetext{b}{Longest axis of SiO emitting region, after deconvolution from the 14$''$ JCMT beam.}
\tablenotetext{c}{The SiO abundance used towards K3-50A for deriving outflow kinematics is the average abundance from the other six sources.}
\label{tab:sio_prop}
\end{deluxetable}
\end{landscape}

\begin{landscape}
\begin{deluxetable}{rccccccccc}
\tablecolumns{10}
\tablewidth{-10pc}
\tablecaption{HCO$^+$ Observations and Derived Infall Properties}
\tablehead{
\colhead{Source} & \colhead{T$_{\rm mb}$} & \colhead{FWZP\tablenotemark{a}} &\colhead{$<\int T_{\rm mv}^*dv>$} & \colhead{$<$log[N$_{\rm HCO^+}$]$>$} &\colhead{$<$V$_{\rm in}$\tablenotemark{b}$>$} & \colhead{$\dot{M}_{\rm in}$} & \colhead{Age} & \colhead{M$_{\rm in}$}&\colhead{M$_{\rm in}$/M$_{\rm out}$}\\
	& \colhead{(K)} &\colhead{(km s$^{-1}$)} & \colhead{(K km s$^{-1}$)} & \colhead{(cm$^{-2}$)} & \colhead{(km s$^{-1}$)} & \colhead{($\times10^{-3}$ M$_\odot$ yr$^{-1}$)} & \colhead{($\times10^4$ yr)} & \colhead{(M$_\odot$)}
}
\startdata
G5.89-0.39	&31.5	&140	&35.6	&13.17	&\nodata&\nodata&0.29&\nodata&\nodata\\		
G10.47+0.03	&11.7	&60	&11.2	&12.67	&1.02$\pm$0.1	&8.89	&0.88	&78.0	&0.42\\
G10.6-0.4	&31.2	&50	&28.4	&13.08	&1.19$\pm$0.1	&27.34	&1.42	&389	&0.78\\
G19.61-0.23	&8.3	&80	&14.4	&12.78	&1.49$\pm$0.6	&$<$8.5&$<$0.71&$<$60.3&$<$0.37\\
G20.08-0.14	&5.9	&55	&9.2	&12.59	&1.35$\pm$0.3	&5.51	&0.60	&33.0	&0.09\\
G29.967-0.02	&16.5	&30	&9.2	&12.59&\nodata&\nodata	&1.19&\nodata&\nodata\\		
K3-50A		&17.0	&30	&13.2	&12.75&\nodata&\nodata	&$<$1.36&\nodata&\nodata\\	
\enddata
\tablenotetext{a}{Full Width at Zero Power of the HCO$^+$ line.}
\tablenotetext{b}{Infall velocity averaged over each pixel with an infall signature}
\label{tab:hco_prop}
\end{deluxetable}
\end{landscape}

\end{document}